\documentclass[fleqn,usenatbib]{mnras}

\usepackage{amsmath}	
\usepackage{amssymb}	
\usepackage{siunitx,url}

\usepackage[dvipdfmx]{graphicx}

\newcommand{\beq}{\begin{equation}}
\newcommand{\beqa}{\begin{eqnarray}}
		  \newcommand{\eeq}{\end{equation}}
\newcommand{\eeqa}{\end{eqnarray}}

\newcommand{\lsim}{\lesssim}
\newcommand{\gsim}{\gtrsim}

\newcommand{\lmk}{\left(}
\newcommand{\rmk}{\right)}
\newcommand{\lnk}{\left\{ }
\newcommand{\rnk}{\right\} }

\newcommand{\lla}{\left\langle}

\newcommand{\rra}{\right\rangle}

\newcommand{\cM}{{\cal M}}
\newcommand{\cL}{{\cal L}}
\newcommand{\cR}{{\cal R}}

\title[]{ Search for Neutron Star Binaries in the Local Group Galaxies Using LISA } 

\begin{document}
\author[N. Seto]{Naoki Seto\\
Department of Physics, Kyoto University, Kyoto 606-8502, Japan
}
\maketitle

\begin{abstract} 
We discuss the prospects of LISA for detecting neutron star binaries (NSBs) in the local group galaxies such as LMC and M31. Using the recently  estimated merger rate   $ {\rm 1540 \, Gpc^{-3}yr^{-1}}$ and inversely applying  the conventional arguments based on the $B$-band galaxy luminosities,  we estimate  the frequency distributions  of NSBs in the local galaxies. We find that, after 10 year observation with its current design sensitivity, LISA might detect $\sim 5$ NSBs  both in LMC and M31 with signal-to-noise ratios larger than 10. Some of the NSBs might be three-dimensionally localized well within LMC. These binaries will be useful for studying various topics  including the origin  of r-process elements.  
\end{abstract}



\section{introduction}



On 17 August 2017, the LIGO-Virgo network detected a gravitational wave (GW) signal GW170817 from a merging NSB (Abbott et al 2017a). Soon after the detection, associated electromagnetic signals have been extensively  observed in various bands and epochs (Abbott et al 2017b). These multi-messenger observations had profound impacts on broad research fields.

In the next 15 years,  GWs from merging NSBs will be steadily detected by ground-based GW detectors, and our knowledge on the related topics will be largely improved. Mentioning just in relation to this paper, the chirp mass distribution of merging NSBs would be well determined, and the characteristics of their host galaxies will be statistically studied, together with  estimations of the  comoving NSB merger rate much  better than the current value $1540^{+3200}_{-1220}{\rm Gpc^{-3} yr^{-1}}$ (Abbott et al. 2017a).  

In this paper, we discuss the prospects of LISA for detecting NSBs in the local galaxies such as the Large Magellanic Cloud (LMC)  and the Andromeda Galaxy (M31).  LISA has optimal sensitivity to  GWs around 0.1-100mHz, and is planned to be launched in 2034 (Amaro-Seoane et al. 2017).  Even though,  with respect to  NSBs,  the detectable range of LISA  is much smaller than that of the current ground-based detectors, LISA can observe a GW from a  NSB well before its merger (e.g., $10^5$\,yr), and thus might be suitable for observing NSBs in the local galaxies.
In contrast, considering the estimated  merger rate, it is statistically improbable that ground based detectors can detect a merging NSB in the local group earlier than the LISA era.

The local group galaxies have played vital roles for development of astronomy, astrophysics and cosmology (van den Bergh 1999). Among others, due to its proximity and size, LMC has been closely investigated in various respects, e.g., its three-dimensional structure,  careful measurements of its distance as an important step of the cosmological distance ladder,  gravitational/hydrodynamical interactions with other galaxies, star formation history, and gravitational lensing effects for detecting dark objects  (see e.g., Alcock et al. 2000; Freedman, et al. 2001;  Alves 2004; van der Marel; Harris \& Zaritsky 2009).  As an irregular galaxy distinct from Milky Way (MW) and M31, LMC  also helps us to more deeply study evolution of galaxies by a comparative method.  
Therefore, even in the LISA era, census survey of NSBs in  the local galaxies will  be valuable.

After GW170817, one of the stimulated  topics  is the origin of r-process elements.  The estimated comoving NSB merger rate and the observed kilonova emissions are considered to be consistent with a scenario that a significant fraction of r-process elements are produced at NSB mergers (e.g., Thielemann et al. 2017; Hotokezaka, Beniamini \& Piran 2018). But, given a large expansion velocity ($\sim 0.1c$) of ejected matter, it would not be straightforward to unambiguously make a spectroscopic confirmation of  r-process elements.
Meanwhile, even with the next generation optical/IR telescopes (e.g., E-ELT, GMT and TMT), it will be still 
 difficult to take high-resolution spectra (Skidmore et al. 2015) for individual stars in the host galaxies of  NSBs detected by ground based detectors.
This is because,   the targets are expected to be too distant. Therefore, the local group survey by LISA would make an interesting contribution also to the studies on the r-process elements.
 In fact,  LMC has been known to have an anomalous Europium profile.   More specifically, its relatively metal rich stars ([Fe/H]$\gsim -1$) show higher Europium abundances [Eu/Fe] than those of MW (Russell \& Bessell 1989; Hill et al. 1995; van der Swaelmen et al. 2013).

 In relation to compact binary search with LISA,  some authors have discussed possibilities  of detecting white dwarf (WD) binaries in the local group galaxies  (e.g., Cooray and Seto 2005; Korol, Koop \& Rossi 2018; Lamberts et al. 2019). 
Since  the chirp masses of NSBs are expected to be narrowly distributed (Tauris et al. 2017; Farrow, Zhu \& Thrane 2019), we can make more solid arguments for detectability of NSBs at given frequencies, compared with WD-WD systems. 
This point would be worth recognising, when discussing scientific cases of LISA.

This paper is organized as follows. In \S 2, we briefly discuss the basic aspects of nearly monochromatic GW emission and estimate the NSB merger rates for the four local group galaxies, LMC, SMC, M31 and M33. We inversely use the conventional arguments based on the $B$-band luminosity of galaxies (Phinney 1991; Kalogera et al. 2001), with the recently estimated NSB merger rate $\rm 1540 Gpc^{-3} yr^{-1}$ (Abbott et al. 2017a). 
In \S 3, we evaluate the total number of NSBs detectable with LISA in each galaxy, and study  their frequency distribution. In \S 4, we discuss the parameter estimation errors for the detectable NSBs, and also mention related astronomical issues. \S 5 is a concise summary of this paper.

\section{neutron star binaries  in local group galaxies}

\subsection{GW emission}

Here we briefly discuss GW emission from a nearly monochromatic circular binary and its orbital evolution. After taking average with respect to the inclination angle (e.g. Robson et al. 2019), the strain amplitude is given by 
\beq
h_{\rm A}=\frac{8G^{5/3}\cM^{5/3}\pi^{2/3}f^{2/3}}{5^{1/2}c^4d} \label{amp}.
\eeq
Here $c$ is the speed of light, $G$ is gravitational constant,  $d$ is distance to the binary, and $f$ is the GW frequency (twice the orbital frequency). The chirp mass $\cM$ is given  by the two masses of the binary as $\cM=(m_1m_2)^{3/5}(m_1+m_2)^{-1/5}$ and we take $\cM=1.2M_\odot$ as a fiducial value (Farrow et al. 2019).

\begin{table*}
\begin{tabular}{l|llllll}
                                   & MW           & LMC                  & SMC                 & M31                 & M33                 \\
\hline
distance [kpc]     & (10) & 50 & 61 & 780 & 840\\
B-band luminosity $L_{B,g}[10^{10}L_\odot]$                     & $0.90 $ & $0.12$     & $0.028$   & $3.5$   & 0.30    \\
NSB merger rate $R_{g}[\rm 10^{-4} yr^{-1}]$ & 1.4 & 0.18 & $0.043$ & $5.4$ & 0.46 \\
$N_g(>2{\rm mHz})$     & 33 & 4.3 & 1.0 & 130 & 11\\
\hline
highest frequency $f_{\rm h}$[mHz] & 7.4 & 3.4& 2.0& 12.3 & 4.9\\
lowest frequency $f_{\rm l}$[mHz] for $T=2\,{\rm yr}$    & (1.6) & 2.6 & 2.7 & NA & NA\\
lowest frequency $f_{\rm l}$[mHz] for  $T=10\,{\rm yr}$    & (1.0) & 1.7 & 1.8 & 6.8 & 7.5\\
\hline
\end{tabular}
\caption{The basic parameters of the nearby galaxies including MW. For MW, we set $d=10$kpc as a reference distance, and use the parenthesis $(\cdots)$ to show the values directly affected by  this setting.  At $T=2$\,yr, eq.(19) has no solution $f_{\rm l}$ for M31 and M33.}
\end{table*}

Below we mainly use the inclination-averaged expression (\ref{amp}) as a characteristic GW amplitude.  To deal with the dependence on the inclination angle  $I$, we supplementarily  use the following expression
\beq
h_{\rm U}(I)\equiv h_{\rm A} F(\cos I).\label{amp2}
\eeq
The factor $F(u)$ is given by   
\beq
F(u)\equiv \sqrt{\frac5{4}} \lnk \frac{(1+u^2)^2}4+u^2  \rnk^{1/2}
\eeq
and normalized as $\lla F(\cos I)^2\rra_I=1$.

For an edge-on binary, we have $F(0)=\sqrt5/4=0.56$.  The  factor becomes  $F(1)=\sqrt{5/2}=1.58$ for a face-on binary. Also, for 66\% of binaries (i.e. $|u|>1/3$), we have $F(u)>0.72$.

Due to gravitational radiation reaction, the GW frequency $f$ changes at the rate
\beqa
 \frac{df}{dt} &=&\frac{96\pi^{8/3}G^{5/3}f^{11/3}\cM^{5/3}}{5c^5}\label{df}\\
& =& {1.3\times 10^{-16}}{\rm s^{-2}} \left(
                                            \frac{f}{{2}{\rm mHz}}
                                           \right)^{11/3} \left(
                                           \frac{\cM}{1.2
                                           M_\odot} \right)^{5/3} .
\eeqa

Note that expressions (\ref{amp}) and (\ref{df}) are given for circular binaries. As the  known Galactic NSBs  will have  evolved eccentricities $e\ll 1$ at $f\gsim 2$mHz (see e.g. Kyutoku et al. 2019), we only discuss NSBs with $e\ll 1$.   For such binaries, eq.(\ref{df})   has a correction factor $(1+O(e^2))$, but this factor would not be important for our arguments below.  Therefore, we use expressions for  circular binaries.

\subsection{merger rate and frequency distribution}

From observed binary pulsars in  MW (and associated globular clusters), the Galactic NSB merger rate $R_{\rm MW}$ has been estimated, taking into account relevant observational effects  (Phinney 1991; Narayan et al. 1991; Kalogera et al. 2001).  Phinney (1991) conservatively extrapolated the comoving NSB merger rate $\cR$ (per volume per time) as
\beq
\cR=\frac{R_{\rm MW} \, \cL_{B}}{L_{B,{\rm MW}}}
\eeq
using the MW $B$-band luminosity $L_{B,{\rm MW}}$ and the comoving $B$-band luminosity density $\cL_B$.    Note that MW is categorized as a spiral galaxy that typically has a higher star formation activity than the old galaxies such as elliptical (E) and lenticular (S0) galaxies.

Now we inversely estimate the NSB merger rate $R_g$ for the four nearby galaxies, LMC, SMC, M31 and M33 ($g$: the label for these galaxies) that are selected on the basis of distances and luminosities. 
We use the following expression
\beq
R_{g}=\frac{\cR\, L_{B,g}}{\cL_{B}}, \label{rg}
\eeq
where $L_{B,g}$ is the $B$-band luminosity of each galaxy.
We use  $\cR = 1540 \, {\rm Gpc^{-3} yr^{-1}}$   (Abbott et al. 2017a) and    $\cL_B=1.0\times 10^8L_\odot {\rm Mpc^{-3}}$ (Kalogera et al. 2001).   As for $L_{B,g}$, we use the observed values presented in Table 1 (de Vaucouleurs et al. 1991).   Our results $R_g$ are shown in Table 1.

Since the host galaxy of the NSB merger  GW170817 is  an S0 galaxy ($B$-band luminosity $\sim 2\times 10^{10}L_\odot$, Fong et al. 2017),  we  did not make a traditional downward correction ($\sim20\%$) of $\cL_B$ for excluding E and S0 galaxies.  Also, to be compatible with the adopted luminosity $L_{B,g}$ of the four galaxies, we ignored an upward correction ($\sim 30\%$) of $\cL_B$ in relation to  the reprocession of blue  light to infrared light by interstellar dust (Phinney 1991; Kalogera et al. 2001).

With the merger rate $R_g$ in hand,   we next estimate the frequency distribution $dN_g/df$ of NSBs in each galaxy, assuming that most of NSBs are formed at frequencies much lower than the regime in our interest $f\gsim1$mHz. Then, applying the continuity equation  in the frequency space, we have
\beq
\frac{dN_g}{df}=R_g \lmk  \frac{df}{dt}\rmk^{-1}\propto f^{-11/3}.
\eeq
After a simple frequency integral, we obtain the cumulative form as
\beqa
 N_g(>f) & =& \frac{5 c^5 R_g}{256 \pi^{8/3} ( G
 \mathcal{M} )^{5/3} f^{8/3}} \\
 & =& 4.3 \left( \frac{\mathcal{M}}{1.2 M_\odot} \right)^{-5/3} \left(
 \frac{f}{{2}{\rm mHz}} \right)^{-8/3}\nonumber \\
& & \times \left(
 \frac{R_g}{\SI{1.8e-5}{yr^{-1}}} \right) .\label{ni}
\eeqa
  In Table 1, we present the numerical results $N_g(>f)$ for $f=2$mHz.
From equation (\ref{ni}), we can also estimate the highest GW frequency $f_{\rm h}$ (omitting the subscript $g$) for NSBs in each galaxy, by solving 
$N_g(>f_{\rm h})=1$.   We present the numerical results $f_{\rm h}$ in Table 1.
 In terms of $f_{\rm h}$, the cumulative function  is simply given by 
\beq
N_g(>f)=\lmk \frac{f}{f_{\rm h}} \rmk^{-8/3}. \label{cu2} 
\eeq

Using eq.(\ref{df}), we can derive the remaining time before the merger as follows
\beq
\int_f^\infty \lmk \frac{df}{dt}\rmk^{-1}df=1.9\times 10^3\,{\rm yr} \lmk \frac{f}{12.3{\rm mHz}} \rmk^{-8/3}  \left(
                                           \frac{\cM}{1.2
                                           M_\odot} \right)^{-5/3} \label{time}
\eeq
with the reference value $f_{\rm h}=12.3$mHz for M31.
 Note that even for $f=f_{\rm h}$,  the remaining time (\ref{time})   (actually given by $R_g^{-1}$) is  much longer than the expected operation period of LISA.   Therefore, our target NSBs are nearly monochromatic. 


\section{NSB detection with LISA}

\subsection{noise spectrum}

We first discuss the noise spectrum $S_{\rm n}(f,T)$ of LISA, averaged over the direction and polarization angles. We decompose the spectrum $S_{\rm n}(f,T)$ into two pieces, the detector noise $S_{\rm d}(f)$ and the confusion noise $S_{\rm c}(f,T)$ as follows
\beq
S_{\rm n}(f,T)=S_{\rm d}(f)+S_{\rm c}(f,T).
\eeq
Here we explicitly show the dependence of the confusion noise $S_{\rm c}(f,T)$ on the operation period $T$,  taking into account the  expected progress of   foreground subtraction. 

For the detector noise $S_{\rm d}(f)$, we use the analytical fitting formula that is given by eq.(3) in Robson et al. (2019) and also shown in Fig.1. This fitting is an excellent approximation, especially at $f\lsim 30$mHz (see their Fig.3). The low frequency part ($f\lsim 1$mHz) is dominated by the acceleration noise and we have asymptotic profile $S_{\rm d}(f)^{1/2}\propto f^{-2}$.     Meanwhile, at $f\gsim 10$mHz, the noise $S_{\rm d}(f)$ is mainly from the position noise, originally from the shot noise. Due to the finiteness of the armlength $L$, the signal cancellation becomes prominent at $f\gsim c/(2\pi L)=$19mHz (for the current design $L=2.5$Gm), and we have the asymptotic profile $S_{\rm d}(f)^{1/2}\propto f^1$ at $f\gg c/(2\pi L)$, as shown in Fig.1.

\begin{figure}
 \includegraphics[width=1\linewidth]{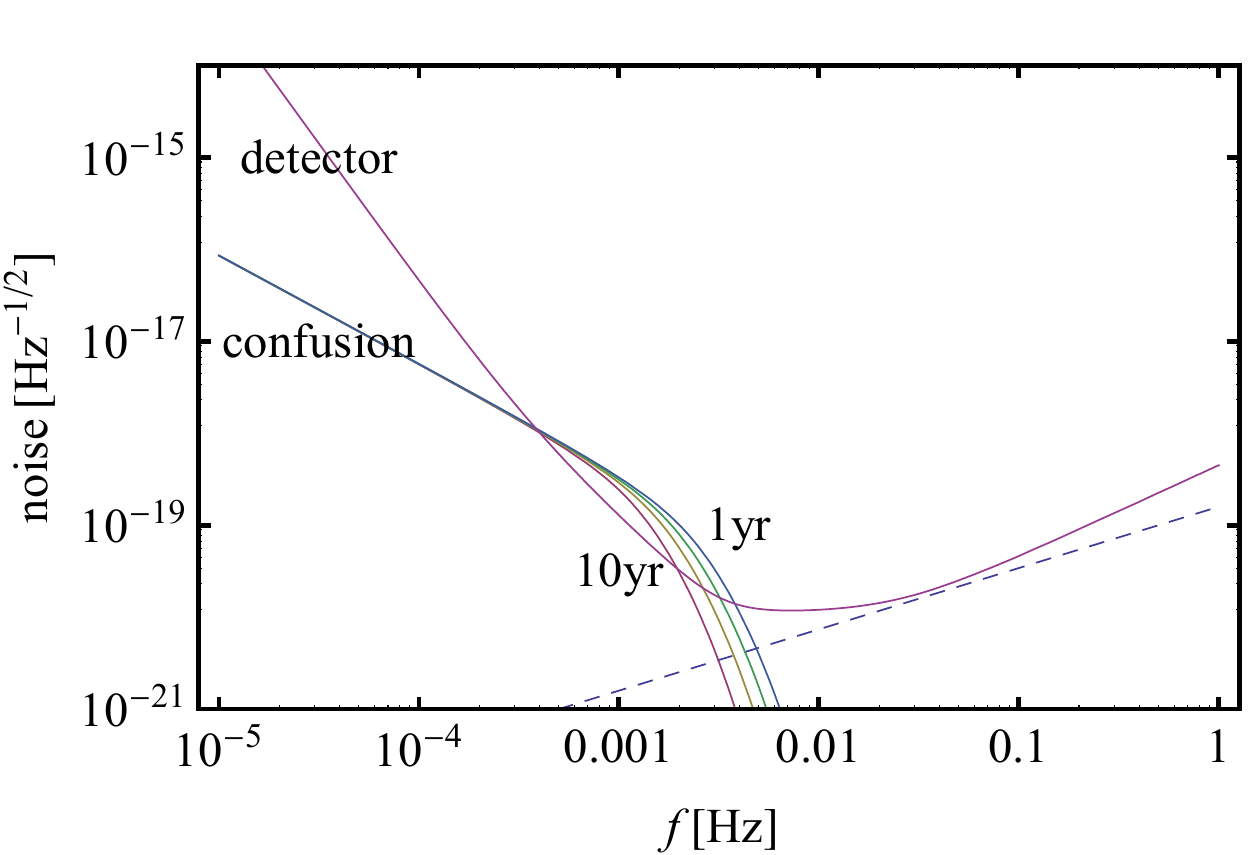} \caption{
The detector noise spectrum $S_{\rm d}(f)^{1/2}$ and confusion noises $S_{\rm c}(f,T)^{1/2}$ for $T=1,2,4$ and 10\,yr (from right to left). The  high frequency part $f\gsim 10$mHz of the detector noise is dominated by the  shot noise and asymptotically becomes $\propto f^1$. The confusion noises are generated  with the scaling relation (\ref{sca}). The dashed line is proportional to $f^{2/3}$ and contacts the detector noise curve at $f_{\rm t}=0.0345$Hz.
} 
\end{figure}

In Fig.1  we also present the confusion noises $S_{\rm c}(f,T)$ for $T=1,2,4$ and 10 yr.  Here we ignored the extra-Galactic component that is expected to be much weaker than the detector noise (Bender \& Hils 1997).  In Fig.1, to incorporate the $T$-dependence of the confusion noise $S_{\rm c}(f,T)$, 
we applied  the following scaling relation (Seto 2002)
\beq
S_{\rm c}(f,T)=\eta^{-7/3} S_{\rm c}(f/\eta,4{\rm yr})\label{sca}
\eeq
 with the conversion factor $\eta\equiv(T/{\rm 4yr})^{-3/11}$.  In this expression, 
the base function  $S_{\rm c}(f,4{\rm yr})$ is given for the specific operation period $T=4$yr, with its actual expression mentioned shortly.
In eq.(\ref{sca}),  the combination $f/\eta$ in the argument originates from the $T$ dependence of the characteristic frequency $f_{\rm c}$ that is associated with the  Galactic foreground subtraction.  The frequency $f_{\rm c}$ is obtained by solving the following equation  \footnote{The actual value of the right-hand side of this equation  is irrelevant for our scaling argument.} for the number of binaries in the unit frequency bin
\beq
\lmk\frac{dN_{\rm MW}}{df}  \rmk_{f_{\rm c}} \frac1T= 1. 
\eeq
On the other hand, the  prefactor $\eta^{7/3}$ in eq.(\ref{sca}) is for recovering the time-independent asymptotic profile 
\beq
S_{\rm c}(f,T)\propto \frac{dN_{\rm MW}}{df} h_{\rm A}^2\propto f^{-7/3}
\eeq
valid for  $f\ll f_c$.

For the actual expression of  the base function $S_{\rm c}(f,4{\rm yr})$, we use  the fitting formula (14) given in Robson et al. (2019) with the updated parameters (Robson private communication)
\[(\alpha,\beta,\kappa,\gamma,f_k)=(0.244, -0.258, 1020, 1145, 0.0011).\]
We confirmed that our scaling relation (\ref{sca}) very accurately reproduces their updated confusion noises curves  $S_{\rm c}(f,T)$ for $T=0.5,1$ and 2\,yr.

As mentions earlier, in this paper, we use the noise spectrum $S_{\rm n}(f,T)$ averaged over the direction and polarization angles.  Compared with short duration signals, this averaging is a much better approximation for a long-term GW observation ($T>1$yr), because of the annual motion of detectors (see e.g. Seto 2004). Furthermore, interestingly, the sky direction of LMC is almost normal ($\sim 85^\circ$) to the ecliptic plane, and polarization angle dependence becomes particularly weak, due to the symmetry of the detector geometry. We can straightforwardly make a more detailed analysis, including direction and polarization dependences religiously. But it would be rather cumbersome and would only produce minor corrections.

\subsection{signal-to-noise ratio}
Next we discuss detection of nearly monochromatic NSBs in the four galaxies. For the observational period $T$, the effective signal strength is given by $h_{\rm A} T^{1/2}\propto f^{2/3} T^{1/2}$. In Fig.1, for each galaxy with a fixed distance $d$ and chirp mass $\cM$, this signal strength is obtained by appropriately shifting the dashed line in the vertical direction with the scaling  $\propto T^{1/2}$. The signal-to-noise ratio $\rho$ of a binary becomes 
\beqa
\rho&=&\frac{h_{\rm A} T^{1/2}}{S_{\rm n}(f,T)^{1/2}}\\
&=&\frac{8G^{5/3}T^{1/2}\cM^{5/3}\pi^{2/3}}{5^{1/2}c^4d} \lmk\frac{f^{2/3}}{S_{\rm n}(f,T)^{1/2}} \rmk .\label{snr}
\eeqa
Below, we set the fiducial value $\rho_{\rm th}=10$ for the detection threshold (see also Moore et al. 2019). 

In eq.(\ref{snr}), the frequency dependence is determined by the combination  $f^{2/3}S_{\rm n}(f,T)^{-1/2}$. For the current design of LISA,  this combination takes its maximum value $5.58\times 10^{-18} {\rm Hz^{7/6}}$ at $f_{\rm t}=34.5$mHz. \footnote{At 34.5mHz, the time before the merger (\ref{time}) is 119\,yr.} Geometrically, as demonstrated in Fig.1, at $f=f_{\rm t}$,  the noise curve $S_{\rm n}(f,t)^{1/2}$ contacts with a curve $\propto f^{2/3}$. 
Around the tangential point $f_{\rm t}$, the combination  $f^{2/3}S_{\rm n}(f,T)^{-1/2}$ weakly depends on frequency $f$.  In fact, in the  frequency interval $\rm [6mHz,~510mHz]$, this combination is within a factor two of the maximum value.

Since we have  $f_{\rm h}\ll f_{\rm t}$ for the four galaxies (see Table 1), the existence of NSBs is negligible  at $f>f_{\rm t}$. In the range $f\lsim f_{\rm h}$ relevant for our study, expression (\ref{snr}) is a monotonically increasing function of $f$ (see also Fig.1).  Therefore, in each galaxy (with fixed $d$ and $\cM$),   to estimate the total number $N_{\rm D}$ of detectable NSBs, we just need to solve the frequency $f=f_{\rm l}$ for the equation 
\beq
\rho=\rho_{\rm th} \label{sol}
\eeq
 with expression (\ref{snr}) for the left-hand side.\footnote{For notational simplicity, we omit the subscript \lq\lq{}$g$\rq\rq{} for $f_{\rm l}$ and $N_{\rm D}$, as in the case of  $f_{\rm h}$.} This frequency $f_{\rm l}$  corresponds to the lowest frequency of detectable NSBs, and does not depend on the  merger rate $R_g$ (for each galaxy), in contrast to the highest one $f_{\rm h}$.
  In Table 1, we present the numerical values of $f_{\rm l}$ for $T=2$ and 10\,yr.  Note that we do not have a solution $f_{\rm l}$ for M31 and M33 at $T=2$\,yr, as explained later with eq.(\ref{r2}). 
If we have a solution $f_{\rm l}$, the total number of detectable NSBs is formally given by eq.(\ref{cu2}) as follows
\beq
N_{\rm D}=N_g(>f_{\rm l})=(f_{\rm l}/f_{\rm h})^{-8/3}.\label{nd}
\eeq

\begin{figure}
 \includegraphics[width=1\linewidth]{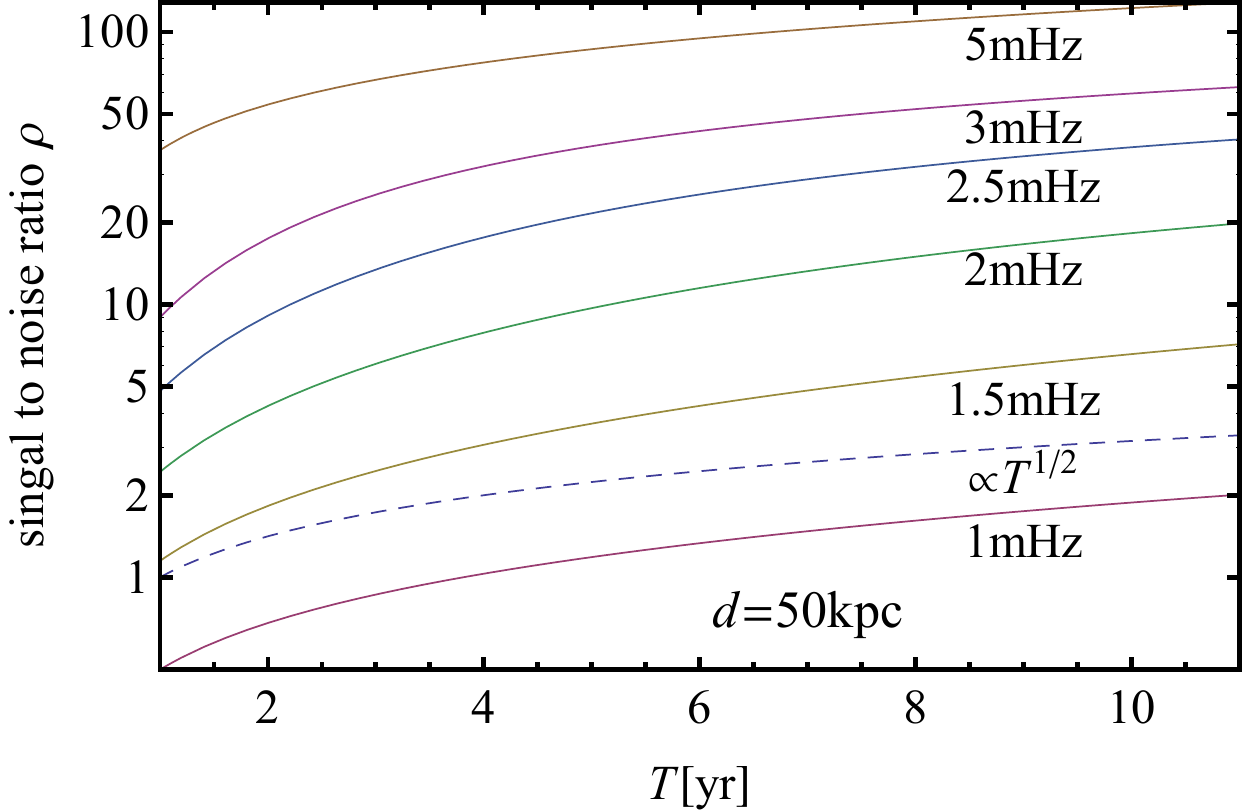} \caption{The signal-to-noise ratios $\rho$ for NSBs  as a function of the observational time $T$. The solid curves are for $f=1,1,5,2,2.5,3$ and 5mHz. The dashed curve is proportional to $T^{1/2}$.  The vertical scaling is  normalized with the distance to LMC 50kpc and the chirp mass $1.2M_\odot$.   } 
\end{figure}

Here we discuss the $T$-dependence of the signal-to-noise ratio $\rho$ given in eq.(\ref{snr}) or more specifically the combination $T^{1/2} S_{\rm n}(f,T)^{-1/2}$. As shown in Fig.1, for a reasonable operation period $1{\rm yr}\lsim T\lsim 10$yr,  the confusion noise show a prominent time variation in the range $\rm [1mHz,5mHz]$.  Above this range, the confusion noise is much smaller than the detector noise and negligible for our study. In Fig.2, we show $\rho \propto  T^{1/2} S_{\rm n}(f,T)^{-1/2}$ with the vertical scale normalized for LMC. As expected, $\rho$ increases more rapidly than $\propto T^{1/2}$ in the range  $\rm [1mHz,5mHz]$.  But we merely have $\rho \propto T^{1/2}$ outside the frequency range.

Now we evaluate the total number of detectable NSBs $N_{\rm D}$ for the four galaxies.  In Fig.3, with solid curves, we show $N_{\rm D}$ for operation period $1{\rm \, yr}<T<20{\rm\, yr}$. Interestingly, we have $N_{\rm D}=0$ at  $T<2.9$\,yr for M31 and similarly at $T<3.4$\,yr for M33. This is because we do not have a solution $f_{\rm l}$ for eq.(\ref{sol}), as mentioned earlier.  
For M31, this can be easily elucidated  with the following expression equivalent to eq.(\ref{snr})
\beqa
\rho&=&10  \lmk  \frac{f^{2/3} S_{\rm n}(f)^{-1/2}}{5.59\times 10^{18} \rm Hz^{7/6}} \rmk              \lmk  \frac{d}{780\rm kpc}  \rmk^{-1}\lmk  \frac{\cM}{1.2M_\odot}  \rmk^{5/3} \nonumber\\
& &\times \lmk  \frac{T}{\rm 2.9yr}  \rmk^{1/2} . \label{r2}
\eeqa
Here the combination $f^{2/3} S_{\rm n}(f)^{-1/2}$ is normalized by its maximum value at $f=f_{\rm t}$ discussed in the previous subsection.  Therefore, even at the optimal frequency $f_{\rm t}$, we need 2.9\,yr to have $\rho=10$.  We can also derive the critical time $2.9\times (840/780)^2=3.4$\,yr for M33, using distances to the two galaxies. 

In the case of M31, if we increase $T$ from 2.9\,yr, the solution $f_{\rm l}$ for eq.(\ref{sol}) swiftly decreases from $f_{\rm t}$, because of the weak frequency dependence of the combination $f^{2/3} S_{\rm n}(f)^{-1/2}$. Then, following eq.(\ref{nd}), the detectable number $N_{\rm D}$ rises rapidly. For $T=10$\,yr, we have the total number  $N_{\rm D}\sim5$, and their signal-to-noise ratios $\rho$ have  a small variation   from $\rho=10$ ($f_{\rm l}=6.8$mHz) to  $\rho=14$ ($f_{\rm h}=14$mHz, see Table 2).  These results for M31 are virtually independent of the confusion noise, as the involved  NSBs have relatively high frequencies.  Fig.3 also shows that for M33, we need $T\gsim 20$\,yr to have $N_{\rm D}\sim 1$, due to  its lower NSB merger rate than that of M31.

\begin{figure}
 \includegraphics[width=1\linewidth]{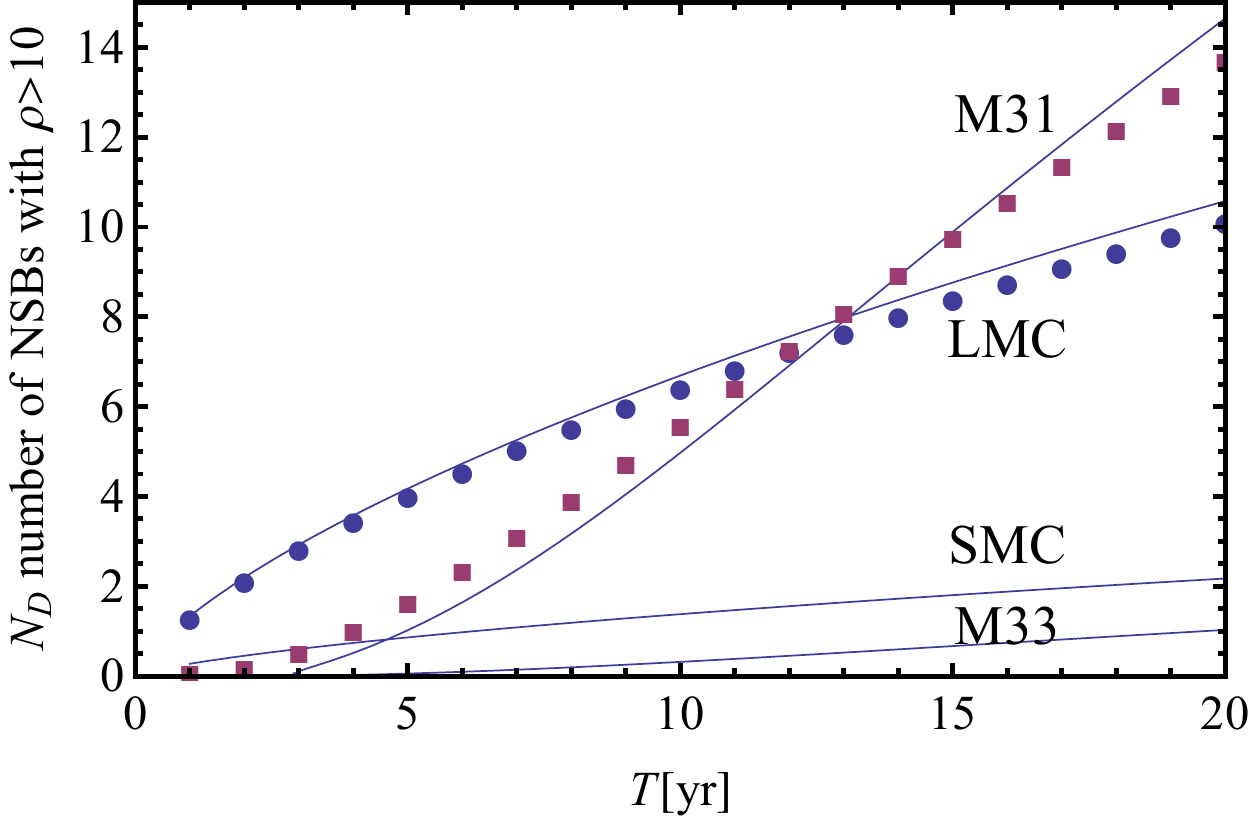} \caption{Number of detectable NSBs with $\rho>10$. For M31 and M33, we have $N_{\rm D}=0$ at $T<2.9$\,yr and $<3.4$\,yr respectively. 
The squares and circles are predictions including the inclination dependence based on  eq.(\ref{amp2}).  }
\end{figure}

\begin{table*}
\begin{center}
\begin{tabular}{c|c|c|c|c|c|c|}
\multicolumn{1}{c|}{ }&
\multicolumn{4}{c|}{LMC}&
\multicolumn{2}{c|}{M31}
\\ \hline
\multicolumn{1}{c|}{{\it T} [yr] }&
\multicolumn{2}{c|}{4}&
\multicolumn{2}{c|}{10}&
\multicolumn{2}{c|}{10}
\\ \hline
frequency type	&$f_{\rm h}$&$f_{\rm l}$&$f_{\rm h}$&$f_{\rm l}$&$f_{\rm h}$&$f_{\rm l}$
\\ \hline
$f$~[mHz] & 3.4& 2.1 & 3.4 & 1.7 & 12.3& 6.8
\\\hline
$\rho$ & 46 & 10 & 76 & 10 & 14& 10
\\\hline
$\Delta \cM/\cM$ & 0.0037 & 0.098 & $3.5\times 10^{-4}$ & 0.037 & $1.8\times 10^{-5}$& $2.3\times 10^{-4}$
\\\hline
$\delta$ [degree] & 0.8 & 5.9 & 0.47 & 7.5 & 0.72& 1.9
\\\hline
$\Delta d/d$ & 0.044 & 0.2 & 0.026 & 0.2 & 0.14& 0.2
\\\hline
\end{tabular}
\end{center}
\caption{Observational prospects for LMC and M31. We present the basic properties of  detectable NSBs at the highest and lowest frequencies. The frequency regime $[f_{\rm l},\,f_{\rm h}]$ for LMC does not overlap with that of M31. }
\end{table*}
 
Meanwhile, in Fig.3, LMC shows a more gradual increase from $N_{\rm D}=1.5$ ($T=1$yr) to 7 (10yr).  The two frequencies $f_{\rm h}$ and $f_{\rm l}$ are both lower than those of M31 (see Table 2), respectively reflecting the lower merger rate $R_g$ of LMC and  its closer distance $d$ (see Table 1). In addition, because of the  complicated shape of the total noise spectrum $S_{\rm n}(f,T)$ in the frequency regime relevant for LMC  (see Fig.1), the signal-to-noise ratios $\rho$  will have a larger scatter. For $T=10$\,yr,  $\rho$ are distributed from    $10$ ($f_{\rm l}=1.7$mHz)  to  $76$ ($f_{\rm h}=3.4$mHz).

Until now, our discussions have been based on the inclination averaged amplitude $h_{\rm A}$ given by eq.(\ref{amp}). But we can easily include the inclination dependence with the correction factor  $F(u)$ in eq.(\ref{amp2}). To estimate the detectable number $N_{\rm D}$ with $F(u)$, we make a segmentation of the variable $u=\cos I$ in the range $[-1,1]$, and count the expected detectable numbers for each segment. Then the  total number $N_{\rm D}$ can be obtained by summing contributions of  all the segments of $u$.  
In Fig.3, for LMC and M31, we show the results with circles and squares. We can see that the simple inclination-averaged scheme (solid curves) 
 captures the basic profiles of the more complicated calculation (circles and triangles).  For M31, at $T\lsim 10$yr, the solid curve underestimate squares. This difference seems reasonable, considering the concentration of  detectable NSBs   around the threshold $\rho_{\rm th}=10$.  The fluctuation of the amplitude by the factor $F(u)$ helps some NSBs exceed the detection threshold. 
In Fig.3, with respect to the solid curve for M31, we had $N_{\rm D}=0$ for $T<2.9$\,yr, as mentioned earlier.  If we include the inclination dependence, the critical operation period  becomes $2.9\times 2/5=1.2$\,yr, given the maximum value $F(1)=\sqrt{5/2}$ for a face-on binary.

So far, we have estimated various quantities for NSBs in the four galaxies. But some of them depend directly on the merger rates $R_g$ that are rather uncertain. Here we  sort out the scaling relations for $R_g$.
For the frequencies, we have
\beq
f_{\rm h}\propto R_g^{3/8},~~f_{\rm l}\propto R_g^0.
\eeq
For the numbers of  NSBs, we have
\beq
\frac{dN_g}{df}\propto R_g,~~N_g(>f)\propto R_g,~~  N_{\rm D}\propto R_g.
\eeq
Even if the $B$-band luminosity $L_{B,g}$ is a good tracer of the NSB merger rate $R_g$ as in eq.(\ref{rg}),  the comoving NSB merger rate  $\cR = 1540^{+3200}_{-1220}{\rm Gpc^{-3} yr^{-1}}$   (Abbott et al. 2017a) currently  contains a large estimation error, corresponding to a correction factor of $3^{\pm 1}$ for Fig.3. As mentioned earlier, our knowledge on the related issues will be continuously improved in the near future.

We have also fixed the detection threshold at $\rho_{\rm th}=10$.  Since the signal-to-noise ratios of M31 are virtually unaffected by the time-dependent  confusion noise, the detectable number $N_{\rm D}$ of M31 (given in Fig.3) has a particularly simple scaling relation.  For a choice $\rho_{\rm th}\ne 10$,  we just need to rescale the time from $T$ to   $T(\rho_{\rm th}/10)^2$ in Fig.3.

We can apply a similar argument to binary black holes (BBHs).  Here we take  $\cM=20M_\odot$ and $\cR=53{\rm Gpc^{-3} yr^{-1}}$ (Abbott et al. 2018) as reference values for  BBHs.  Then, for each of the four galaxies, the highest frequency $f_{\rm h}$ becomes $(1540/53)^{-3/8} (20/1.2)^{-5/3}=0.05$ times smaller than the result shown in Table 1. As a consequence, even with $T=10$\,yr, the maximum signal-to-nose ratios at $f=f_{\rm h}$ are 81, 1.6 and 3.6 respectively for MW, LMC and M31.  It might be difficult  to observe  a BBH in the local group galaxies other than MW (Seto 2016)

\section{parameter estimation errors}

  Now we discuss how accurately  we can estimate  the basic parameters of  NSBs  detected in the local galaxies.  Given the expected number $N_{\rm D}$ in Fig.3,   we concentrate on LMC and M31.

First, we briefly summarize the parameter estimation errors, following Takahashi \& Seto (2002).  They used the Fisher matrix analysis for totally eight fitting parameters, including relevant angular variables.  In relation to our studies on NSBs, we quote their numerical results valid for $T\gsim$2\,yr and $f\gsim 1$mHz. The error for the frequency derivative is given by 
\beqa
\Delta {\dot f} &\simeq& 0.43 \lmk \frac{\rho}{10} \rmk^{-1} T^{-2}\\
&=&4.3 \times 10^{-18} {\rm s^{-2}}  \lmk \frac{\rho}{10} \rmk^{-1} \lmk  \frac{T}{\rm 10yr}\rmk ^{-2}.
\eeqa
For a circular binary, from the relation ${\dot f}\propto \cM^{5/3}f^{11/3}$, we can evaluate the relative error for the chirp mass (more precisely  the redshifted one)  as follows
\beq
 \frac{\Delta \cM}{\cM}\simeq\frac35\frac{\Delta {\dot f}}{\dot f}.\label{dm}
\eeq

Meanwhile, the typical magnitude for the amplitude error is given by
\beq
\frac{\Delta h_{\rm A}}{h_{\rm A}}\simeq0.2 \lmk \frac\rho{10}  \rmk^{-1}.
\eeq
For $\Delta \cM/\cM \ll \Delta h_{\rm A}/h_{\rm A}  \ll 1$ (approximately valid for  the cases below), we have the  distance error
\beq
\frac{\Delta d}{d}\simeq \frac{\Delta h_{\rm A}}{h_{\rm A}}.
\eeq

To characterize the angular resolution of a binary in the sky, we define $\delta$ as the square root of area of the error ellipse.   Its typical magnitude is given  by
\beq
\delta \simeq 1.25^\circ   \lmk \frac{f}{10 \rm mHz}  \rmk^{-1} \lmk \frac\rho{10}  \rmk^{-1}. \label{ang}
\eeq
Here we used eq.(15) in Takahashi \& Seto (2002) for the characteristic area of the error ellipse. 
Note that for $f\gsim 1$mHz, the source direction is mainly determined by the frequency modulation caused by the revolution of LISA  around the Sun (Cutler 1998).

In Table 2, we provide examples of $\Delta \cM/\cM, \delta $ and $\Delta d/d$ for NSBs at the highest and lowest frequencies.  In this table, we simply put $  {\Delta \cM}/{\cM}\simeq3/5 \times {\Delta {\dot f}}/{\dot f}$ given in eq.(\ref{dm}) for circular binaries. 
Because of the strong frequency dependence ${\dot f}\propto f^{11/3}$ of the denominator in eq.(\ref{dm}),   M31 generally has smaller  ratios $\Delta {\cM}/{\ \cM} $ than LMC.    But, considering the potential effects such as the residual eccentricity and peculiar velocity,  the correspondence (\ref{dm})  will not be valid for a small error (e.g., $\Delta \cM/\cM \lsim 10^{-3}$). In addition,  such precision might not be fully utilized for astronomical studies, considering the intrinsic scatter $\sim \Delta \cM/\cM\sim 0.1$ of observed binary pulsars (Forrow et al. 2019).

At LISA data analysis,  the measured chirp mass $\cM$ would be the primary information for selecting a NSB candidate. Given the aforementioned scatter of observed chirp masses, the measurement error of a level $\Delta \cM/\cM\lsim 0.05$ would not become  the primary  obstacle for the selection. 
One of the potential concerns for a secure identification of a NSB is a confusion with  NS-WD and WD-WD systems that have  relatively large  chirp masses $\cM\simeq1.2M_\odot$ (like the NS-WD system PSR B2303+46).  At present, we cannot make a solid statement about the outlook of this discrimination.   But there is a possibility that the numbers of such high mass  systems might be small (for WD-WD systems see e.g., Nelemans et al. 2001; Korol et al. 2017; Lamberts et al. 2019). In addition,  we can expect that multi-messenger observation (LISA+SKA, LSST, and so on)  would significantly improve our knowledge on related issues such as the chirp mass distribution of Galactic WD-WDs and NS-WDs (e.g., Littenberg et al. 2013; Korol et al. 2017; Kyutoku et al. 2019). 

Next we discuss the accuracy of the  three-dimensional position estimation for  detected NSBs. LMC has the distance $d=50$kpc and radius $\sim 5$kpc.  Therefore, its angular size is $\sim 10^\circ$ with its line-of-sight width $\sim 0.1$. 
These numbers should be compared with $(\delta, \Delta d/d)$ in Table 2. 
For the NSB at the detection limit $f=f_{\rm l}$, we will be just able to confirm its association to LMC.  In contrast, for the best one at $f=f_{\rm h}$,  the good localization will allow us to make interesting  astronomical studies (e.g., calibration of the cosmic distance ladder).

Meanwhile, M31 has the mean distance $d=780$kpc and the characteristic size  $\sim 30$kpc, corresponding to the angular size $\sim 5^\circ$ with the radial width 0.04.  As shown  in Table 2, the distance error $\Delta d/d$ is much larger than the radial width.  But  we can roughly localize the detected NSBs in the sky image of M31.  Here, we should notice that, for $f=f_{\rm h}$, the error $\delta$ for M31 is comparable to that for LMC,  even though the signal-to-noise ratios $\rho$ are largely different (see Table 2).  This is because the sky direction is determined  by  using the frequency modulation, and  is more efficient for higher frequency sources, as indicated by eq.(\ref{ang}).

\section{Summary}

In the next 15 years, ground-based detectors will observe a large number of  NSB mergers. However,  it is statistically improbable that  one of them is in the local group. To search for GWs from a NSB in the local group, we need to observe  a lower frequency  signal well before the merger. If detected,  because of its proximity, the binary would play interesting roles for astrophysical studies, including the origin  of r-process elements.

In this paper, we have discussed the prospects of NSB observation with LISA for the representative local galaxies such as  LMC and M31.  We inversely applied the conventional argument based on $B$-band galactic luminosities, and estimated the expected NSB merger rates $\sim 2\times 10^{-5} {\rm yr^{-1}}$ for LMC and $\sim 5\times 10^{-4} {\rm yr^{-1}}$ for M31. Here we used the  recently  estimated merger rate   $ {\rm 1540 Gpc^{-3}yr^{-1}}$ that has a uncertainty of $\times3^{\pm1}$ (Abbott et al. 2017a).   Then we showed that, for the estimated merger rates,  LISA can detect $\sim 5$ NSBs both in LMC and M31 in 10 years.    More conservatively speaking, even if the actual merger rates are 5 times smaller than our \lq\lq{$B$-band}\rq\rq{} estimation, we will detect $\sim 1$ NSB in the two galaxies. 

As shown in Fig.3, the number of detected NSBs would depend strongly  on the observational period. 
Compared with detected NSBs in M31,  those in LMC  would have lower frequencies (see Table 2). Some of the detected NSBs could be localized in the sky with an accuracy of $\sim 0.5^\circ$,  well in the sky images of the two galaxies. 
 For the highest frequency  NSB in the LMC, we might determine its distance within a few percent error, and might use it also for calibrating the cosmic distance ladder.

\section*{Acknowledgments}

The author would like to thank N. Cornish, K. Kyutoku, T. Narikawa, T. Robson, H. Sugiura and D. Toyouchi for their help. 
This  work is supported by JSPS Kakenhi Grants-in-Aid for
Scientific Research (Nos. 17H06358 and 19K03870).



\end{document}